# Effect of pressure on the transport properties and thermoelectric performance of Dirac semimetal ZrTe$_5$

Sanskar Mishra[1], Nagendra Singh[1], V.K. Gangwar[2], Rajan Walia[3], Manindra Kumar[1], Udai Bhan Singh[1], Deepash Sekhar Saini[1], Jianping Sun[4], Genfu Chen[4], Dilip Bhoi[5], Sandip Chatterjee[6], Yoshiya Uwatoko[5], Jinguang Cheng[4*], Prashant Shahi[1†]

[1]*Department of Physics, DDU Gorakhpur University, Gorakhpur,273009, India*
[2]*Department of Physics, K.G.K. (P.G.) College, Moradabad, 244001, India*
[3]*Department of Physics, University of Allahabad, Prayagraj, 211002, India*
[4]*Beijing National Laboratory for Condensed Matter Physics and Institute of Physics, Chinese Academy of Sciences, Beijing, 100190 China*
[5]*Institute for Solid State Physics, University of Tokyo, Kashiwa, Chiba 277-8581, Japan*
[6]*Department of Physics, Indian institute of technology (IIT)-BHU, Varanasi, 221005, India*

**ABSTRACT**

ZrTe$_5$ has been extensively studied due to its novel topological properties, which are tunable by subjecting to various parameters like temperature, chemical synthesis conditions, and pressure. In this study, we have investigated and compared the effect of hydrostatic pressure up to ~20 kbar on the transport properties of ZrTe$_5$ single crystals grown by chemical vapor transport (CVT) and flux methods. With the application of pressure, the electrical resistivity $\rho(T)$ and thermopower $S(T)$ of both crystals were found to increase in the whole temperature range unlike the other known thermoelectric materials, such as Bi$_2$Te$_3$, SnSe etc. This observation is supported by the complementary first-principles band structure calculation as the application of pressure widens the direct bandgap at Γ point. Moreover, the analysis of the pressure dependent magneto-transport and Shubnikov de-Hass oscillation results revealed an increase in carrier concentration and effective mass along with the reduction of mobility as pressure rises. Furthermore, with the application of pressure, the flux-grown ZrTe$_5$ crystals display a transition from unipolar to bipolar charge transport as evidenced by the emergence of resistivity peak at $T^*$ under high pressure, unlike the CVT-grown ZrTe$_5$ crystals where the bipolar charge transport near its characteristic resistivity peak ($T_p$) remains unaffected. Our study also reveals pressure-induced enhancement of $T_p$ and $T^*$ for both crystals, suggesting an upward shift of Fermi level upon compression. Additionally, for the CVT-grown ZrTe$_5$ crystals, the application of pressure nearly doubled the thermoelectric power factor (*PF*) at 18.2 kbar and room temperature. In contrast, for the flux-grown ZrTe$_5$ crystals, the *PF* varies weakly as the pressure is raised to 17 kbar. Our results underscore the role of crystals synthesis technique along with the application of pressure as an effective strategy to optimize the magneto transport and thermoelectric performance of ZrTe$_5$.

## I. INTRODUCTION

ZrTe$_5$, a 2D layered material with relatively weak interlayer bonding (~12 meV/Å$^2$)[1], has been studied for a long time owing to its anomalous transport properties [2–7]. This material is widely recognised due to a Lifshitz transition [8] at a characteristic temperature ($T_p$) , as evidenced by a characteristic peak in resistivity, sign reversal of Seebeck coefficient and Hall resistivity [3]. Angle resolved photoemission spectroscopy (ARPES) studies revealed the shift of chemical potential across the band gap from hole-like bands to the electron-like bands across $T_p$ as temperature is lowered from room temperature [9]. More interestingly, several works [10, 11] demonstrated that the shift in chemical potential is sensitively affected by the Te concentration in ZrTe$_5$, resulting in a variation of $T_p$ as well as the transport properties.

†prashant.phy@ddugu.ac.in; *jgcheng@iphy.ac.cn

Therefore, it is likely that the samples used by different groups have different stochiometric proportions of Te [11]. This is evident by the difference in the transport properties between the crystals grown by chemical vapor transport (CVT) and flux methods. For the CVT grown crystals, the $T_p$ tends to be higher and varies in between ~130-150 K [12–15], whereas the flux grown crystals usually display $T_p$ lower than ~60 K [9,15–17]. In addition, depending on the band shift [12], the anomalous Hall effect (AHE) and chiral magnetic effect are reported in $ZrTe_5$ crystals synthesized by the flux method [18,19].

Besides Te vacancies, the physical properties and the Fermi surface topology of $ZrTe_5$ also vary sensitively with application of pressure [17,20–22]. Zhou *et al*. reported a complete suppression of resistivity peak in $ZrTe_5$ around 6.2 GPa along with the emergence of superconductivity below a transition temperature $T_c$ ~ 2.0 K, which is found to increase with pressure, reaching as high as ~6 K at 21.2 GPa [23]. Similar results related to pressure-induced superconductivity has also been reported in $HfTe_5$ [24]. Investigations in the low-pressure regime below 2 GPa revealed more interesting results, including pressure-induced topological phase transitions and changes in Fermi surface topology. J. L. Zhang *et al*. reported a non-trivial to trivial Berry phase shift around ~2 GPa in the CVT grown $ZrTe_5$ crystals along with the large suppression of magnetoresistance (MR) as pressure is applied. This has been attributed to the pressure induced change in the Dirac fermion state near the Fermi level [25]. First-principles calculation under pressure reveals a transition from strong to weak topological insulator phase mediated by an intermediate Dirac semi-metallic phase [25]. D. Santos-Cottin *et al*. [26] measured the transport and optical properties of both the CVT and flux grown crystals. In contrast to the behaviour reported in Ref. [27], $T_p$ was found to increase with pressure. On the other hand, $T_p$ in a-few-layer-thick $ZrTe_5$ nanoflakes does not change with pressure [20]. Similarly, $T_p$ of the sister compound $HfTe_5$ almost remains flat up to 1 GPa and thereafter keeps decreasing [22]. These results suggest that the variation of $T_p$ that depends on the sample synthesis conditions and varies differently with the application of pressure remains poorly understood in a unified picture.

Therefore, in this study, we investigated the physical properties of $ZrTe_5$ single crystals under the hydrostatic pressures ($0 \leq P \leq 20$ kbar) using the magneto-transport, thermopower ($S$) and complimentary band structure calculations. A comparison between the $ZrTe_5$ single crystals synthesized by CVT and flux techniques revealed a huge difference in the pressure induced transport properties. In addition, we observed that both resistivity ($\rho$) and $S$ increases with increasing pressure, irrespective of the crystal growth techniques, suggesting an unconventional nature in reference of other layered thermoelectric materials [28–30]. Notably, the resistivity peak at both $T_p$ and $T^*$ of both crystals displays a dramatic variation with pressure. In the CVT crystals, $T_p$ increases up to 155 K around 13 kbar and followed by a slight decrease upon further increasing pressure. In contrast, a peak-like feature at $T^*$ emerges in $\rho(T)$ of flux-grown crystals with a moderate application of pressure ~6 kbar and gradually shifts towards the higher temperatures. Hall resistivity $\rho_{xy}(B)$ under pressure of CVT samples is qualitatively same as that of the ambient pressure, while there is deviation for the flux samples. At low temperatures, the CVT crystals display a very pronounced quantum (SdH) oscillation at ambient and high pressures, whereas no such features could be observed for the flux crystals. This points out the significant difference in Fermi surface topology of $ZrTe_5$ crystals prepared by different methods. Analyses of magneto-transport phenomenon under pressure revealed an increase in overall carrier density and effective mass with a concurrent decrease of carrier mobility for both crystals. Furthermore, band structure calculations reveal the Dirac like band structure near the Fermi level at ambient pressure, as pressure increases the band gap also increases. In addition, we also observed a pressure-induced enhancement in thermoelectric power factor ($PF = S^2/\rho$). Our results suggest that the pressure application is an effective strategy to optimize $PF$ and reveals $ZrTe_5$ as a promising candidate for energy conversion technology.

## II. EXPERIMENTAL AND COMPUTATIONAL DETAILS

The $ZrTe_5$ single crystals were grown by the CVT and flux method as described in Refs. [31,32]. Details about the characterizations of these single crystals have been reported in our previous study [10]. Resistivity was measured using a conventional four-probe method, with the current flowing along the longest direction of crystal, *i.e.* the *a* axis. The temperature-dependent $S(T)$ at ambient and high pressures was measured using a homemade setup (described in an earlier work of our group [30]) integrated with a self-clamped piston cylinder cell (PCC). For the $S(T)$ measurements, the temperature gradient of $\Delta T/T \sim 1\%$ was maintained along the *a* axis with the help of a thin-film heater (resistance ~ 120 Ω). A AuFe/Chromel differential thermocouple was used to measure temperature difference ($\Delta T$) with a precision of ~ 0.2%. The maximum error in the measurement of $S(T)$ is ~ 0.5 $\mu V/K$. For high-pressure measurements, Daphne 7373 was used as the pressure transmitting medium. A manganin wire resistance gauge was used to probe the pressure inside the pressure cell. For MR and Hall resistance measurements cubic anvil high pressure cell was used, in which current injected along the *a* axis and magnetic field is applied along the *b* axis.

The density functional theory calculations for the structural optimizations, density of states (DOS) and electronic band structure were performed using the Quantum ESPRESSO package [33]. Exchange-correlation effects were treated with the generalized gradient approximation (GGA) using the Perdew-Burke-Ernzerhof (PBE) functional [34] along with Grimme's D3 dispersion corrections for van der Waals interactions [35]. The force and total energy convergence thresholds were set to $10^{-5}$ atomic units and $10^{-12}$ Ry, respectively, with a wavefunction (WFC) cut off 80 Ry. Structural optimizations were performed at both ambient and high-pressure conditions using a pressure convergence threshold of 0.05 kbar. For these calculations, scalar-relativistic ultra-soft pseudopotentials were employed. DOS and band structure calculations were performed using fully relativistic pseudopotentials from the PSLibrary [36], taking spin-orbit coupling (SOC) into account. For the band structure calculations, 15×15×4 k-point mesh was used, and a denser k-point grid (double of the original mesh) was applied for the non-self-consistent field DOS calculations to accurately map the Fermi level.

## III. RESULTS
### A. Transport properties at ambient pressure

ZrTe$_5$ crystallizes in an orthorhombic layered crystal structure with space group *Cmcm*. Figure 1(a) depicts the crystal structure of ZrTe$_5$, in which the trigonal prismatic chains of ZrTe$_3$ runs along the *a* axis and are connected with each other by zig-zag Te atoms along the *c* axis, forming a two-dimensional layer of ZrTe$_5$ in the *ac* planes. These two-dimensional layers of ZrTe$_5$ are stacked together via weak van der Waals interaction to form a 3D ZrTe$_5$ crystal. The interlayer bonding energy is as low as graphite (~12 meV/Å$^2$), making it suitable for 2D cleaving [1]. Figure 1(b) shows the image of ZrTe$_5$ single crystals grown by the CVT and flux method. Both crystals are featured in needle-like morphology with the longest direction along the crystallographic *a* axis. The CVT crystals are relatively larger in comparison with the flux crystals. Figure 1(c) displays the temperature dependence of *a*-axis resistivity $\rho(T)$ of both single crystals, showing dramatic different behaviours, i.e., the CVT crystal shows a pronounced peak in $\rho(T)$ at $T_P$ ~ 131 K, while the $\rho(T)$ of flux crystals remains semiconducting throughout the measured temperature range. In Fig. 1(d), we have plotted the temperature dependence of thermopower $S(T)$ for both crystals. Like the resistivity, the $S(T)$ also shows distinct behaviours for these two crystals. The $S(T)$ of the CVT crystal displays a sign change at $T_P$ from the high-temperature *p*-type to low-temperature *n*-type dominated charge carriers, while the $S(T)$ of the flux crystal remains positive in the entire investigated temperature range (2-300K). It is important to note that the values of thermopower at room temperature for both crystals are similar ~200 $\frac{\mu V}{K}$, which is relatively high among known thermoelectric materials.

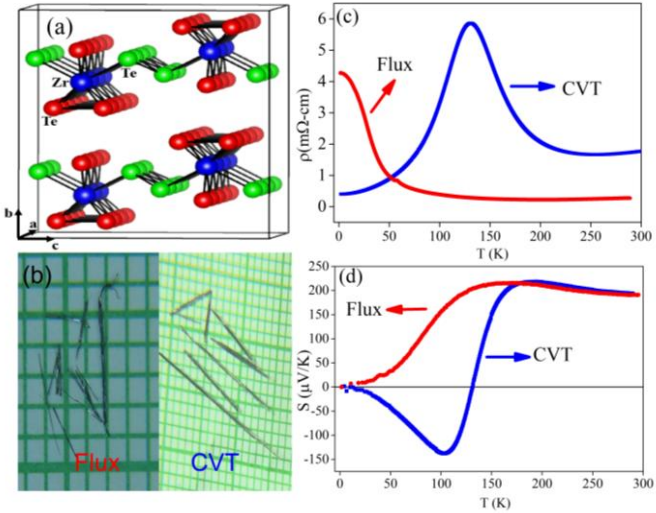

**Figure 1.** (a) Crystal Structure of ZrTe$_5$, in which the trigonal prismatic chains of ZrTe$_3$ are running along the *a* axis and two chains are connected by Zig-Zag Te atoms. Two dimensional layers of ZrTe$_5$ are stacked along the *b* axis. (b) CVT and flux grown single crystals of ZrTe$_5$ with the longest dimension up to ~ 5 mm. (c, d) Temperature dependence of *a*-axis resistivity $\rho(T)$ and thermopower $S(T)$ of ZrTe$_5$ crystals grown by the CVT and flux method.

### B. Transport properties at high pressures

To investigate the effect of pressure on transport properties of ZrTe$_5$, we measured $\rho(T)$ and $S(T)$ of both crystals at different pressures up to 20 kbar in the temperature range $2 < T < 300$ K. Figure 2(a) shows the $\rho(T)$ of the CVT grown crystals under different pressures up to 18.2 kbar. We found that the resistivity exhibits a dramatic enhancement with increasing pressure throughout the entire temperature range, except at low temperatures below 50 K. Notably, the magnitude of $\rho$ at $T_p$ rises significantly, reaching as high as five times at 18.2 kbar. In addition, $T_p$ varies non-monotonically with pressure, *i.e.*, it first increases with pressure up to 12 kbar and then decrease slightly at 18.2 kbar. Near the room temperature, a slight increment of $\rho$ is also observed though it is less pronounced compared at $T_p$. Figure 2(b) shows the $S(T)$ of the CVT grown crystal at different pressures. At all pressures, $S(T)$ changes sign at $T_p$ from positive to negative upon cooling down from room temperature. In line with the resistivity results in Fig. 2(a), the magnitude of thermopower peaks on both sides of $T_p$ also increases with pressure, especially that below $T_p$ is nearly triply enhanced. In addition, the magnitude of $S$ at room temperature increases continuously with increasing pressure, revealing an ~86% enhancement at 18.2 kbar compared to that at ambient pressure.

For comparison, we plotted the $\rho(T)$ and $S(T)$ of the flux-grown ZrTe$_5$ crystals at different pressures in Fig. 2(c) and (d), respectively. As seen in Fig. 2(c), $\rho(T)$ is significantly enhanced by pressure, especially at low-temperature region. In addition, a hump-like feature appears at $T^*$ in $\rho(T)$ at pressures above 4.7 kbar and this feature becomes more evident and moves towards higher temperatures in a fashion similar to the resistivity peak in the CVT crystals. Accordingly, $S(T)$ of the flux-grown sample also exhibits dramatic changes. The $S(T)$ remains positive in the whole investigated temperature and pressure range, indicating the dominant hole carriers. In concomitant with the emergence of resistivity hump, a shoulder-like feature appears near $T^*$ in $S(T)$, moves to higher temperatures and becomes more evident with increasing pressure. $S$ attains maxima at ~150 K and its magnitude increases significantly from 200 to 350 µV/K as pressure increases from ambient to 17 kbar. In addition, the $S$ is enhanced by more than 50% in a broad temperature range between 150 and 290 K.

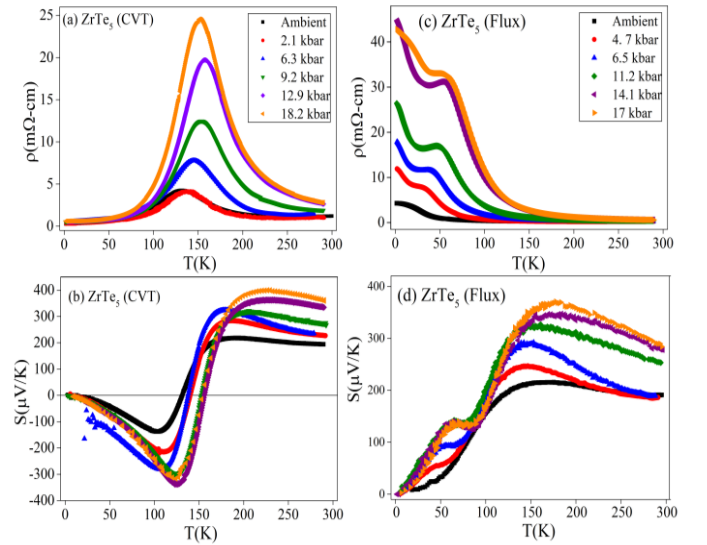

**Figure 2**. Temperature dependences of resistivity and thermopower for (a, b) the CVT and (c, d) the flux grown ZrTe$_5$ crystals at different pressures.

### C. Magneto-transport properties at high pressures

Previously, we have studied the longitudinal MR and Hall resistivity ($\rho_{xy}$) for both CVT and flux crystals at ambient pressure [10]. For the CVT sample at ambient pressure, MR at 1.8 K reaches up to ~900% at 5 T, decreases to about 100% at 100 K and then increases to ~400% up to 170 K upon increasing temperature. Also, the MR at $T \leq 10$ K exhibits pronounced SdH quantum oscillations. In accordance with $S(T)$, the $\rho_{xy}(H)$ shows negative slope at $T < T_p$ due to the dominance of electron carriers and changes to positive slope at all $T > T_p$. At $T_p$, the $\rho_{xy}(H)$ at lower field shows a positive slope due to the presence of high mobility hole carriers and it then changes to a negative slope because of the high-density electron carriers. In contrast to the CVT samples, the MR of the flux samples exhibits a saturation behaviour at ambient pressure. The maximum value of MR ~760 % at 5 T was achieved at 100 K. At all temperatures, the $\rho_{xy}(H)$ exhibit positive slope due to the dominated hole carriers, in line with the $S(T)$ data.

To understand the impact of pressure on the physical properties observed above, we measured MR and $\rho_{xy}$ for both CVT and flux-grown crystals at few selected temperature and pressures using cubic anvil high pressure cell. Figure 3(a, b) and (c, d) show the MR and $\rho_{xy}$ of the CVT crystals at 10 and 20 kbar, respectively. For both samples, the MR is supressed dramatically as the pressure increases. For the CVT sample at 1.5 K and 5 T, the MR~ 55% at 10 kbar decreases to ~25% at 20 kbar. The maximum value of MR ~70% at 5T takes place at 155K under 10 kbar, while it decreases to ~50% at 110 K under 20 kbar. For the flux crystals, in comparison with the MR at 1.8 K and 5 T under ambient pressure, the corresponding MR value has been suppressed to roughly half at 10 kbar and nearly one forth at 20 kbar. The maximum value of MR ~450% at 5 T is found to take place at 100 K under 10 kbar and it decreases to ~200 % at 20 kbar. These observations of pressure induced suppression of MR is consistent with the previous report [23,27]. MR at low temperature also exhibits power-law dependence of magnetic field under pressure, MR $\propto H^n$, where n ~1.5(0.8) for the CVT (flux) crystals. For $\rho_{xy}$, the magnetic field dependence at high pressure is quite similar to that at ambient pressure, except that the magnitude is diminished gradually with pressure.

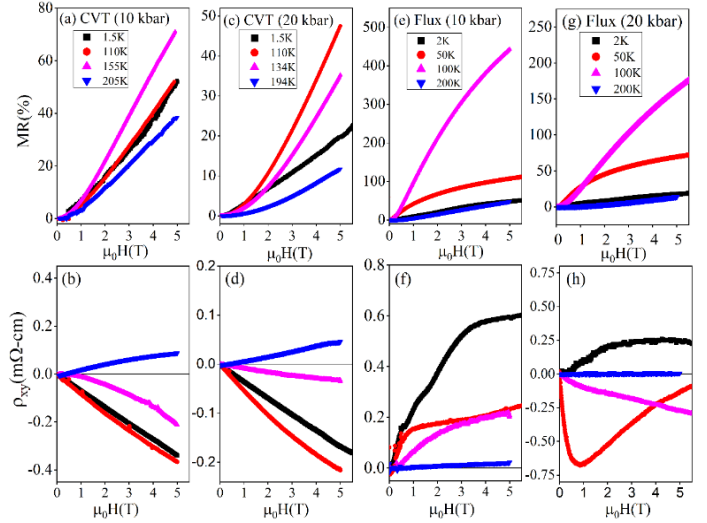

**Figure 3**. Field dependence of MR and $\rho_{xy}$ at (a-d)10 kbar and (e-h) 20 kbar of the CVT and flux-grown ZrTe$_5$ crystals.

To obtain more insight into the evolution of transport behaviours, we employed a two-carrier model to fit the Hall conductivity, $\sigma_{xy} = \frac{\rho_{xy}^2}{\rho_{xx}^2 + \rho_{xy}^2}$, as done previously for the data at ambient pressure, viz.,

$$\sigma_{xy}(B) = eB(\pm \frac{n_1 \mu_1^2}{(1+\mu_1^2 B^2)} \pm \frac{n_2 \mu_2^2}{(1+\mu_2^2 B^2)}) \quad (1)$$

Here, +(-)$n$ stands for hole(electron) density, $\mu$ is mobility of carriers. Figure 4 illustrates the carrier density and mobility for the CVT and flux grown crystals at different pressures. For comparison, we have also plotted the ambient-pressure $n$ and $\mu$ from our previous study [10]. Figure 4(a) presents the $n$ and $\mu$ of the CVT sample at ambient pressure taken from our previous study [10]. The solid symbol stands for electron and open symbol represents hole carries. As can be seen, two-type electron carriers, e$_1$ and e$_2$, are present at $T < T_p$, while one electron (e$_1$) and one hole (h$_1$) carrier with high mobility coexist at $T_p$. The density of hole increases at $T > T_p$. Upon applying pressure on the CVT crystals, the carrier density is moderately increased, while the mobility is reduced roughly to around one third of the value at ambient pressure. The vertical dash lines in Fig. 4(a-c) represent the location of $T_p$ for the CVT grown ZrTe$_5$ single crystals.

For the flux samples, the carrier density at ambient pressure is about two orders lower than that of the CVT samples due to the difference of Te concentrations. As the temperature is raised at ambient pressure, the carrier density increases while the mobility only changes moderately. Upon applying pressure, the carriers' density keeps increasing and reaches as high as ~$10^{27}$ m$^{-3}$ while the mobility decreases to as low as ~$0.2 \frac{m^2}{Vs}$. Additionally, the application of high pressure creates an electron-type carrier in the flux crystals as seen directly from the $\rho_{xy}(H)$ data at 20 kbar, Fig. 3(h). These data reveal that the application of high pressure produces a notable change in the electronic band structures across the investigated pressure range. The vertical dash lines in Fig. 4(e, f) depict the $T^*$ for the flux crystals determined above.

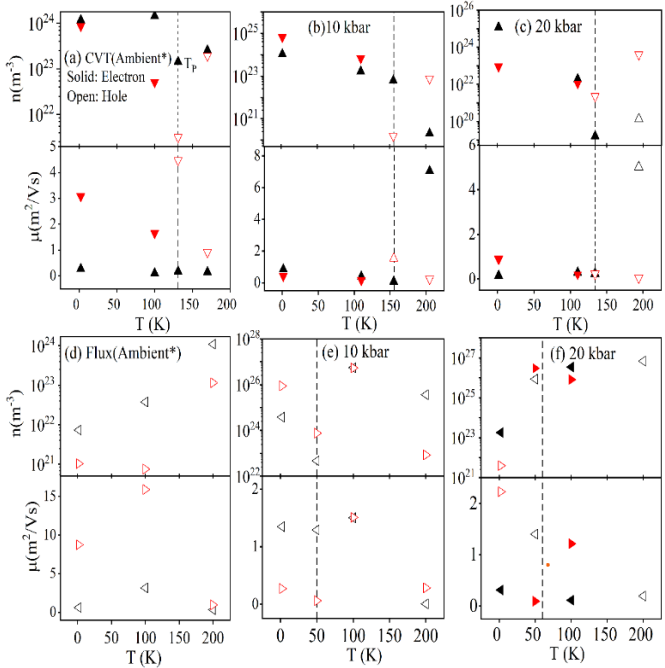

**Figure 4**. Carrier density and mobility of (a-c) the CVT and (d-f) the flux crystals at ambient, 10 and 20 kbar, respectively. For comparison, we have taken the ambient-pressure data from our previous study [10]. Vertical dashed lines depict the location of $T_p$ and $T^*$ for the CVT and Flux crystals respectively.

### D. Shubnikov de-Haas (SdH) quantum oscillation under pressure

One key difference between the CVT and flux grown crystals is manifested in the SdH oscillations. For the CVT crystals, quantum oscillations are clearly visible in the field dependence of MR at ambient and high pressures. In contrast, quantum oscillations are absent in the flux-grown crystals. Therefore, we investigated the SdH oscillations in the CVT grown crystals at selected pressures. Figure 5(a-c) present the background subtracted $\Delta\rho_{xx}$ extracted from the longitudinal MR as a function of $1/B$ at ambient pressure, 20 and 30 kbar, respectively. The corresponding fast Fourier transformation (FFT) at each pressure are shown in the inset. As the pressure increases, the oscillations start to appear at higher magnetic field and the corresponding FFT reveals an increase of the frequency suggesting the increase in the FS area. The analysis of the temperature-dependent FFT amplitude revealed an increase in the effective mass. The FS cross sectional area calculated using the Onsager relation, $S_F = \frac{2\pi e}{\hbar}\alpha$ (where $\alpha$ is the FFT frequency) increases by about six times from $3.99 \times 10^{-4}$ Å$^{-2}$ at ambient pressure to $2.69 \times 10^{-3}$ Å$^{-2}$ at 30 kbar. Similarly, the $m^*$ also increases by four times.

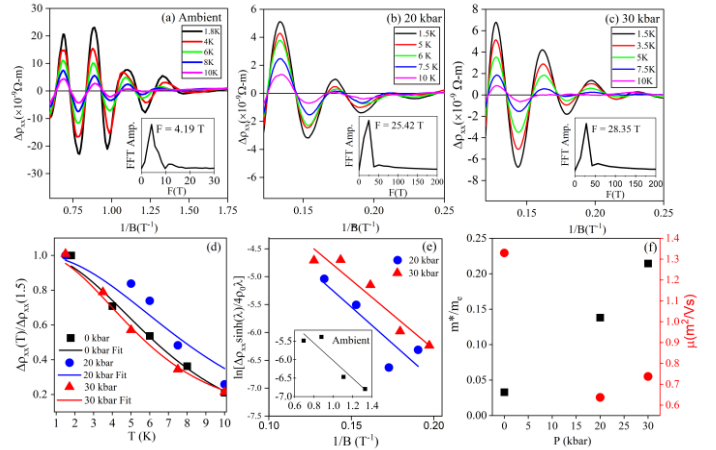

**Figure 5**. (a-c) The SdH oscillations of the CVT samples below 10 K at ambient, 20 and 30 kbar respectively. The inset shows the amplitude vs. frequency of fast Fourier transformation (FFT) of corresponding oscillations. (d) Temperature-dependent damping of normalised amplitude of oscillations at ambient, 20 and 30 kbar. The solid curves represent the fitting of $\Delta\rho_{xx}(T)/\Delta\rho_{xx}(T_{\min})$ vs. $T$ according to the Lifshitz-Kosevich theory as described in Equation (3). (e) The Dingle plot at 20 and 30 kbar. The solid straight line shows the fitting according to Equation (4). The inset of (e) shows the corresponding data for ambient pressure. (f) The calculated cyclotron mass ($m^*$) and quantum mobility of carries at different pressures.

To estimate different parameters related to FS, we analysed the SdH oscillations according to the Lifshitz-Kosevich (LK) theory [8,37,38], which describes the reduction of SdH oscillation amplitude with increasing temperature by the formula, *viz.*,

$$\Delta\rho_{xx} \propto R_T R_D R_S \cos\left(2\pi\left(\frac{\alpha}{B}+\beta\right)\right), \quad (2)$$

where $R_T$, $R_D$ and $R_S$ are three reduction factors responsible for smearing out the oscillations due to temperature, scattering and spin-induced splitting, respectively. Temperature dependence of the oscillation is described by

$$R_T = \frac{\lambda(T)}{\sinh\lambda(T)}, \quad (3)$$

where $\lambda(T) = \frac{2\pi^2 m_* k_B T}{\hbar eB}$ and $m^*$ is the cyclotron mass. Figure 5(d) shows the fitting of Eq. (2) to the $\Delta\rho_{xx}(T)/\Delta\rho_{xx}$ vs. $T$ at ambient, 20 and 30 kbar. Quantum lifetime of carries contributing to SdH oscillation can be determined experimentally using the Dingle Plot [39,40]. According to Lifshitz-Kosevich theory the amplitude of SdH oscillation decreases due to scattering as given by

$$R_D = \frac{\Delta\rho_{xx}}{\rho_0} = \frac{4\lambda e^{-\lambda_D}}{\sinh(\lambda)}, \quad (4)$$

where $\rho_0$ is the resistivity of ZrTe$_5$ in zero magnetic field, and $\lambda_D = \frac{2\pi^2 m_* K_B T_D}{\hbar eB}$ and $\lambda = \frac{2\pi^2 K_B T m_*}{\hbar eB}$. Here, $T_D = \frac{\hbar}{2\pi K_B \tau}$ is called Dingle temperature and $\tau$ is total scattering time of carriers. Figure 5(e) shows a graph of $\ln\left[\frac{\Delta\rho_{xx}\sinh(\lambda)}{4\rho_0\lambda}\right]$ vs $1/B$ at 20 and 30 kbar. The inset shows the same plot for the ambient pressure. The estimated cyclotron mass ($m^*$) and quantum mobility ($\mu$) is summarised in Fig. 5(f). It is clear that $m^*$ increases almost four times as pressure increases to 30 kbar. At the same time, the quantum mobility is reduced to almost half. The analysis of SdH oscillations for the CVT samples are consistent with the results obtained from the two-band model fitting to the magneto-transport data as discussed in previous section.

### E. Computation Results: Band structure and density of States

To obtain further insight into the underlying phenomena, we performed density functional theory band structure calculations. Figure 6(a) displays the Brillouin zone of ZrTe$_5$ centred at Γ point, which highlights the high symmetry points used for band structure calculations. Figure 6(b) illustrate the calculated total density of states (DOS) near the Fermi level at three pressures, *i.e.*, ambient, 10 and 20 kbar. The DOS profile provides the insight into the pressure induced changes in electronic structure near the Fermi level. In Fig. 6(c-e), we present the calculated electronic band structure of ZrTe$_5$ under ambient, 10, and 20 kbar, respectively. At ambient pressure, the conduction band minimum (CBM) and the valance band maximum (CBM) touches exactly at the Γ- point revealing the formation of gapless Dirac like band dispersion near the Fermi level, consistent with previous experimental observations [5,13,15,16,41,42]. This feature is also visible by the presence of a finite DOS near the Fermi level at ambient pressure. As pressure is increased to 10 kbar, the system seems to undergo a phase transition with a noticeable increase in band gap ~70 meV. The bandgap continue to widens further to 90 meV at 20 kbar in agreement with the previous report [1,25].

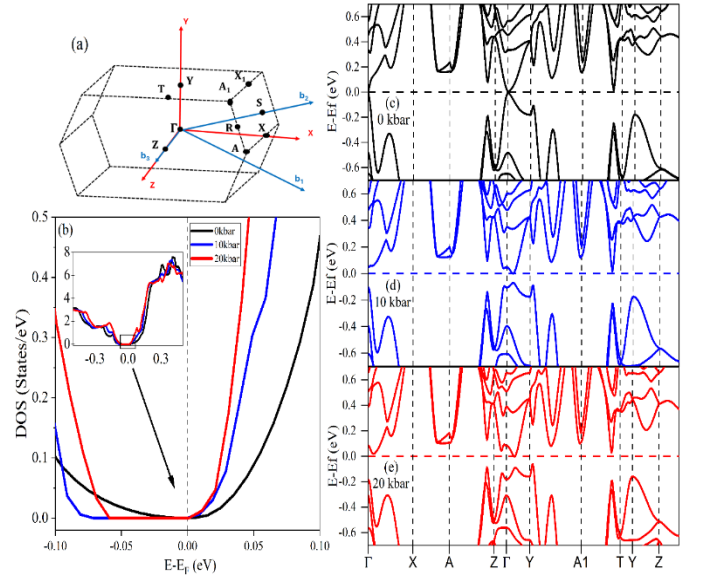

**Figure 6**. (a) The Brillouin zone of ZrTe$_5$ along with high symmetry points used in calculations of bands. (b) Total density of states per eV of ZrTe$_5$ near the Fermi level at different pressures, ambient, 10 and 20 kbar. (c-e) The calculated electronic band structures at ambient, 10 and 20 kbar, respectively.

### IV. DISCUSSION

The difference in the transport properties between the CVT and flux-grown ZrTe$_5$ crystals at ambient pressure arises from the different Te content. In general, the CVT grown crystals are Te deficient, which acts as n-type dopant leading to higher carrier density [Fig.4 (a,d)]. Furthermore, in CVT grown crystals the dominant carrier changes from hole to electron across $T_p$, while near $T_p$ both type of carriers participates simultaneously in transport enabling the bi-polar conduction, whereas the flux grown crystals are close to stoichiometry, hence displaying intrinsic unipolar *p*-type conduction [10]. Despite this difference, the application of external pressure results in a similar overall enhancement of $\rho$ and $S$ for both type of crystals. Figure 7 summarises the pressure dependence of $\rho$ at selected temperature (*2K, Tp/T\* and 290K*) for CVT (Fig.7(a)) and flux (Fig.7(b)) grown crystals. In figure 7(c) we show pressure dependence of $S$ at *290K* for both crystals while Fig7(d) shows the variation of $T_p$ and $T^*$. The $\rho$ enhancement is more pronounced at $T_p$ for the CVT crystal and at low temperatures for the flux crystals as shown in Fig. 7(a, b). For the flux crystals, the $\rho$ remains semiconducting up to 18 kbar, but a broad hump-like feature at $T^*$ appear in $\rho(T)$ above a moderate pressure (4.7kbar) [Fig. 2(c)]. Like $T_p$, $T^*$ gradually shifts towards higher temperatures with increasing pressure [Fig.7(d)]. Variation of $S$ with pressure near room temperature is almost similar in both type of crystals [Fig.7(c)]. However, unlike the CVT grown crystals, the $S$ does not change sign at $T^*$ in the flux grown samples, rather displays a shoulder. The sign change of $S$ and Hall coefficient at $T_p$ occurs due to the compensation of the electron and hole carriers. The absence of sign changes in the flux grown crystals suggests that the carriers are not exactly compensated, but electron like carriers also contribute to the charge transport in addition to the hole type. This is also evident from the sign change of $\rho_{xy}$ around 50-100 K at 20 kbar as in Fig. 3(h). These observations indicate that with the application of pressure the charge conduction change from uni-polar to bi-polar nature in flux grown crystals. This is further bolstered by appearance of peak-like feature in $R(T)$ at 40 kbar in the flux crystals (Appendix Figure 1). In general, $T_p$ in the ZrTe5 single crystals is associated with the Fermi level, which depends on the carrier concentration [26]. Analyses of the magneto-transport properties reveal qualitatively similar impact of pressure on both crystals, *i.e.*, the increase of carrier concentration, resulting the enhancement of $T_p$ and $T^*$ with pressure.

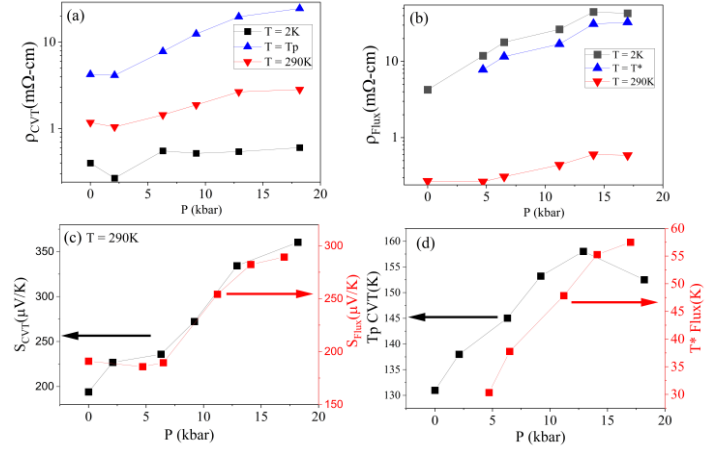

**Figure 7**. Variation of resistivity with pressure at some selected temperatures 2 K, $T_p$ and 290 K for (a) CVT and (b) flux crystals. Pressure dependences of thermopower at 290 K and $T_p$ for (c) the CVT and (d) flux crystals.

In Fig. 8, we have shown thermoelectric power factor ($PF \equiv S^2/\rho$) of the CVT and flux grown ZrTe5 crystals at different pressures. The *PF* of the CVT crystals displays a non-monotonic variation as a function of pressure in the entire temperature region. In addition, it is approximately two times higher in the semiconducting region ($T > T_P$) than in the metallic region ($T < T_P$) at all pressures. In contrast, the *PF* of flux-grown crystal exhibits an almost decreasing trend with pressure. At room temperature, the decrease in *PF* is moderate, while below the room temperature, the *PF* reduction is dramatic, decreasing to almost half near 150 K. Furthermore, on applying a moderate pressure, the room temperature *PF* of the CVT grown ZrTe5 crystals exhibit a dramatic enhancement of ~56.5% at 18.2 kbar. On the other hand, the flux grown crystals show only a weak increase of 6.2% at 17 kbar pressure. Apart from the room temperature, it is also worthwhile to discuss the behaviour of *PF*, below and above $T_p$ (or $T^*$). $S$ shows dramatic enhancement around $T_p$ or $T^*$, however due to the sign change of $S$ across $T_p$ in the CVT crystals the *PF* diminishes at $T_p$. But above and below $T_p$, *PF* is almost enhanced by a factor of 2. On the other hand, the *PF* of the flux grown crystals decreases to 1/3 of that at ambient pressure.

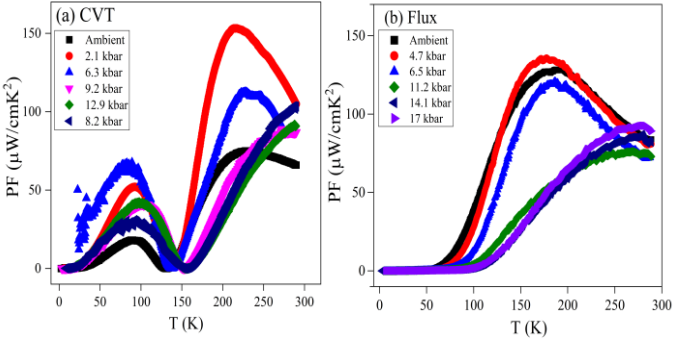

**Figure 8.** Thermoelectric power factor as function of temperature of the (a) CVT and (b) flux grown crystals at different pressure.

Engineering of carrier density and mobility can play a crucial role in optimization of the thermoelectric performance of a material [43–45]. In general, the relationship for $\rho(T)$ and $S(T)$ in presence of electrons and holes can be given by [46]
$\rho(T) = \frac{1}{e(n_e\mu_e + n_h\mu_h)}$, $S = \frac{k_B}{e} \cdot \frac{n_h A^h - n_e A^n}{n_h + n_e}$, where $n_{e(h)}$, $\mu_{e(h)}$ are the density and mobility of electrons(holes), $k_B$ is the Boltzmann constant, $e$ is electronic charge, $A^{h(e)}$ represents average contribution in thermopower from holes (electrons). The pressure induced interplay between density and mobility causes the enhancement in resistivity and thermopower. On the other hand, the observed decrease in mobility with pressure suggests that carries are becoming heavier, indicating the enhancement of effective mass under pressure. This increase in effective mass could be the reason for the significant increase in thermopower. In fact, this expectation is consistent with the quantum oscillation of $ZrTe_5$ under pressure as shown in [27] and the theoretical calculations in Refs. [21,27]. However, the resistivity increase and low temperature upturn in the flux grown crystals with application pressure should be attributed to the band gap opening under compression as indicated by our DFT calculation for stochiometric $ZrTe_5$ crystal, in good agreement with previous studies [48,21]. These results demonstrate that the promising potential of $ZrTe_5$ as a candidate for thermoelectric energy conversion. Moreover, this will pave the way to explore the thermoelectric performance of $ZrTe_5$ in the two-dimensional limit.

## V. CONCLUSION

The objective of this study was to investigate the effect of high pressure on transport properties of the CVT and flux grown $ZrTe_5$ single crystals. We find that the application of pressure has significantly changed the transport properties of $ZrTe_5$ depending on the growth method. While the CVT grown $ZrTe_5$ displays a bipolar transport near $T_p$, the flux crystals also show a bipolar transport character under pressures. The Te vacancies in the CVT crystals can be considered as the form of chemical pressure, however the emergence of bipolar transport characteristic in flux grown crystals marked by emergence of resistivity peak $(T^*)$ resembles that of CVT crystals which highlights the equivalence of physical and chemical pressure effects. Transport measurement along with the complementary first-principles band structure calculations demonstrated the pressure induced opening of direct band gap at Γ point with decreased DOS near the fermi level. The analysis of the magneto transport results at various pressures revealed an increases of carrier concentration and decrease in carrier mobility coupled with an increase in effective mass. Additionally, the application of pressure has been found to enhances the *PF* of CVT grown $ZrTe_5$ single crystals more than 50%, whereas the *PF* enhancement is moderate ~6% for the flux grown crystal at room temperature. Our findings offer valuable insights into the effect of different growth methodologies, and the application of high pressure on the Dirac semimetal $ZrTe_5$. These results open new avenue for engineering thermoelectric materials with enhanced efficiency.

## ACKNOWLEDGEMENT
S.M. acknowledges the support from the Science and Engineering Research Board (SERB), India, for the Junior Research Fellowship (File No.- SRG/2019/001187). V.K.G. acknowledges to the SERB India for the award of the SERB International Research Experience (SIRE) fellowship (File No.- SIR/2022/000804). P.S. acknowledges to the SERB India for the award of the Start-up research grant (SRG/2019/001187), University Grant Commission (UGC) for Basic Scientific Research (BSR) fund (UGC File No.- 30-505/2020(BSR)), UGC-DAE Centre for providing the fund (CRS/2021-22/01/415) and SERB India for SERB-SIRE fellowship (File No.- SIR/2022/000752). J.G.C. is supported by the National Key Research and Development Program of China (2023YFA1406100, 2021YFA1400200), the National Natural Science Foundation of China

## APPENDIX

### 1. ZrTe$_5$ (Flux) data at 40 kbar pressure

We have measured the resistance $R(T)$ and Hall resistance $R_{xy}$ ($T$ = 2 K) of the flux grown crystals at 40 kbar in cubic anvil cell. It exhibits similar behaviours as that of the CVT grown crystals.

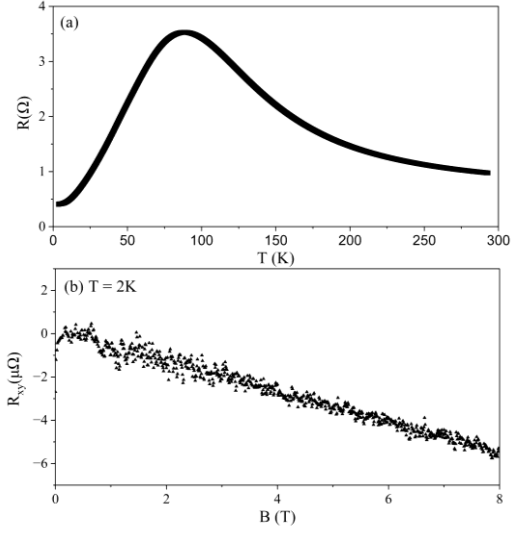

**Appendix figure 1**. The longitudinal and Hall resistance of the flux-grown ZrTe$_5$ crystals at 40 kbar.

### 2. Berry Phase calculation

For estimating the change in Berry phase with pressure, we have applied Lifshitz-Onsager quantization rule [47–49] to analyse the quantum oscillations. According to this rule, each peak and valley of the quantum oscillation can be indexed by Landua level ($n$) as a function of $1/B$. Appendix Fig. 2 depicts the plot of Landau level peak index ($n$) vs. the inverse of the magnetic field ($1/B$) at 0, 20 and 30 kbar pressure. For the calculation of peak index ($n$), we multiplied the values of $1/B$ at each peak and valley of oscillation by FFT frequency. Linear fitting of Fan diagram gives,

$$\frac{\alpha}{B} = n + \beta$$

Where, α is the slope of the Fan diagram which provides details about Fermi surface. The intercept β gives the information about Berry Phase of the system. The intercept $\beta$ can be expressed in terms of Berry Phase as-

$$\beta = \frac{1}{2} + \delta - \frac{\phi_B}{2\pi}$$

where $\phi_B$ is the Berry Phase shift. Here, $\delta$ is an additional phase arises due to degree of dimensionality of the materials. For 2D materials $\delta = 0$, whereas for the 3D materials its value lies between $+\frac{1}{8}$ to $-\frac{1}{8}$. If the value of intercept $|\beta|$ is $0 \pm \frac{1}{8}$, In this case $\phi_B = \pi$ is called trivial Berry Phase. When its value is $\frac{1}{2} \pm \frac{1}{8}$, this situation is called non-trivial Berry Phase ($\phi_B = 0$) [50,51]. In present case, form the linear fit of Landau level fan diagram the value of $\beta \sim 0$ at ambient pressure therefore $\phi_B = \pi$, which gives the non-trivial berry phase and application of pressure gives $\beta \sim 0.5$ which gives $\phi_B = 0$ and therefore trivial berry phase under pressure. Therefore, we see a pressure induced Berry phase shift from non-trivial to trivial.

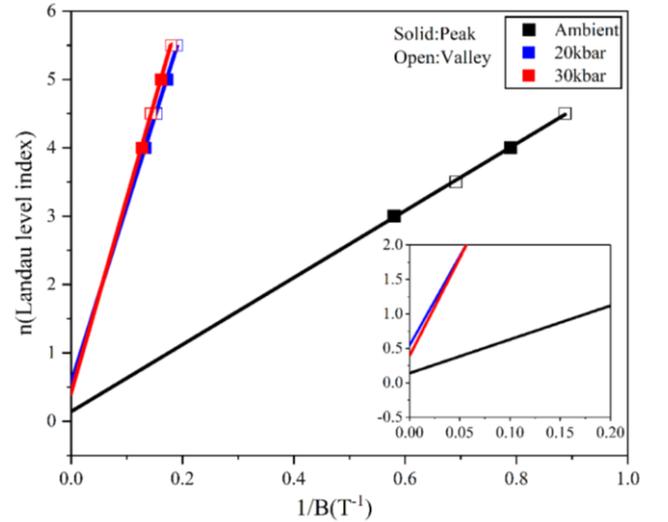

**Appendix Figure 2**. Landau level index ($n$) vs. $1/B$ at different pressures. The inset diagram shows the closer view of intercepts at $1/B = 0$.

### 3. Band structure and Density of States with Experimental lattice constants

We have also calculated the band structure and density of states (DOS) using experimental lattice constant instead of optimized one reported in Ref. [10]. In this case the band gap at gamma point is nearly 178 meV.

11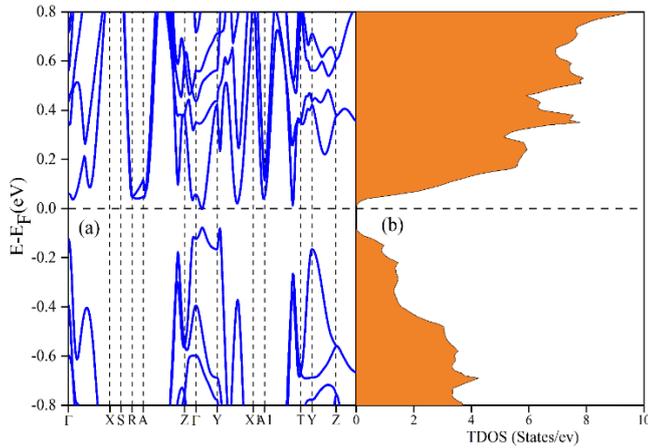

**Appendix Figure 3**. Band structure and density of states with experimental lattice constant at ambient pressure.

# REFERENCES

[1] H. Weng, X. Dai, and Z. Fang, Transition-metal pentatelluride ZrTe5 and HfTe5: A paradigm for large-gap quantum spin hall insulators, Phys. Rev. X **4**, 1 (2014).

[2] S. Okada, T. Sambongi, and M. Ido, Giant Resistivity Anomaly in ZrTe5, J. Phys. Soc. Jpn. 49, 839 (1980).

[3] R. T. Littleton, T. M. Tritt, J. W. Kolis, and D. R. Ketchum, Transition-metal pentatellurides as potential low-temperature thermoelectric refrigeration materials, Phys. Rev. B - Condens. Matter Mater. Phys. **60**, 13453 (1999).

[4] M. Rubinstein, HfTe5 and ZrTe5: Possible polaronic conductors, Phys. Rev. B - Condens. Matter Mater. Phys. **60**, 1627 (1999).

[5] Q. Li, D. E. Kharzeev, C. Zhang, Y. Huang, I. Pletikosić, A. V. Fedorov, R. D. Zhong, J. A. Schneeloch, G. D. Gu, and T. Valla, Chiral magnetic effect in ZrTe 5, Nat. Phys. **12**, 550 (2016).

[6] L. Moreschini, J. C. Johannsen, H. Berger, J. Denlinger, C. Jozwiack, E. Rotenberg, K. S. Kim, A. Bostwick, and M. Grioni, Nature and topology of the low-energy states in ZrTe5, Phys. Rev. B **94**, 1 (2016).

[7] G. N. Kamm, D. J. Gillespie, A. C. Ehrlich, T. J. Wieting, and F. Levy, Fermi surface, effective masses, and Dingle temperatures of ZrTe5 as derived from the Shubnikovde Haas effect, Phys. Rev. B **31**, 7617 (1985).

[8] H. Chi, C. Zhang, G. Gu, D. E. Kharzeev, X. Dai, and Q. Li, Lifshitz transition mediated electronic transport anomaly in bulk ZrTe5, New J. Phys. **19**, (2017).

[9] Y. Zhang, C. Wang, L. Yu, G. Liu, A. Liang, J. Huang, S. Nie, X. Sun, Y. Zhang, B. Shen, J. Liu, H. Weng, L. Zhao, G. Chen, X. Jia, C. Hu, Y. Ding, W. Zhao, Q. Gao, C. Li, S. He, L. Zhao, F. Zhang, S. Zhang, F. Yang, Z. Wang, Q. Peng, X. Dai, Z. Fang, Z. Xu, C. Chen, and X. J. Zhou, Electronic evidence of temperature-induced Lifshitz transition and topological nature in ZrTe5, Nat. Commun. 8, 15512 (2017).

[10] P. Shahi, D. J. Singh, J. P. Sun, L. X. Zhao, G. F. Chen, Y. Y. Lv, J. Li, J. Q. Yan, D. G. Mandrus, and J. G. Cheng, Bipolar Conduction as the Possible Origin of the Electronic Transition in Pentatellurides: Metallic vs Semiconducting Behavior, Phys. Rev. X **8**, 1 (2018).

[11] W. Song, X. Yuan, J. Shen, Y. Xu, F. Wang, X. Zhu, and Y. Zhang, Temperature dependence of band shifts induced by impurity ionization in ZrTe5, Phys. Rev. B **106**, 115124 (2022).

[12] G. Manzoni, A. Sterzi, A. Crepaldi, M. Diego, F. Cilento, M. Zacchigna, Ph. Bugnon, H. Berger, A. Magrez, M. Grioni, and F. Parmigiani, Ultrafast optical control of the electronic properties of ZrTe5, Phys. Rev. Lett. **115**, 207402 (2015).

[13] G. Zheng, J. Lu, X. Zhu, W. Ning, Y. Han, H. Zhang, J. Zhang, C. Xi, J. Yang, H. Du, K. Yang, Y. Zhang, and M. Tian, Transport evidence for the three-dimensional Dirac semimetal phase in ZrTe5, Phys. Rev. B **93**, 115414 (2016).

[14] G. Manzoni, L. Gragnaniello, G. Autès, T. Kuhn, A. Sterzi, F. Cilento, M. Zacchigna, V. Enenkel, I. Vobornik, L. Barba, F. Bisti, Ph. Bugnon, A. Magrez, V. N. Strocov, H. Berger, O. V. Yazyev, M. Fonin, F. Parmigiani, and A. Crepaldi, Evidence for a strong topological insulator phase in ZrTe5, Phys. Rev. Lett. **117**, 237601 (2016).

[15] E. Martino, I. Crassee, G. Eguchi, D. Santos-Cottin, R. D. Zhong, G. D. Gu, H. Berger, Z. Rukelj, M. Orlita, C. C. Homes,